\documentstyle[twocolumn,prc,aps]{revtex}
\input epsf
\begin{document}
\twocolumn[\hsize\textwidth\columnwidth\hsize\csname @twocolumnfalse\endcsname 
\title{Covariance of Antiproton Yield and Source Size in
Nuclear Collisions}

\author{Sean Gavin and Claude Pruneau}
\address{
Department of Physics and Astronomy, Wayne State University, 
Detroit, MI, 48202}
\date{\today} 
\maketitle
\begin{abstract}
We confront for the first time the widely-held belief that combined
event-by-event information from quark gluon plasma signals can
reduce the ambiguity of the individual signals. We illustrate
specifically how the measured antiproton yield combined with the
information from pion-pion HBT correlations can be used to
identify novel event classes.

\vspace{0.1in}
\end{abstract}

]

\begin{narrowtext}

\section{introduction}
 
Signals of novel phenomena from hadron production in relativistic
heavy ion experiments have proven ambiguous at the Brookhaven AGS and
the CERN SPS, primarily because descriptions of the `ordinary' hadron
production mechanisms are under-constrained. It has long been
conjectured -- but not yet demonstrated -- that the added information
from the correlation of distinct signals can reduce the ambiguity of
the individual signals. In particular, the measurement of such
correlations is one of the driving principles behind the PHENIX and
STAR experiments at RHIC \cite{rhic}.  However, to describe the
correlated information that these experiments yield, phenomenologists
must introduce new unconstrained parameters.

We propose that added information can in fact be gained by studying 
antiproton production in conjunction with pion interferometry, HBT, on
an event-by-event basis.  Various authors have suggested that both
antiproton production \cite{qgpPbar} and HBT radius parameters
\cite{PrattBertsch} can change abruptly and dramatically if quark
gluon plasma forms in a heavy ion collision.  However, theoretical
uncertainties \cite{pratt,HeinzJacak,pbarTheory} and experimental
difficulties \cite{pbarData} have made results difficult to interpret
in that context.  In this paper, we demonstrate the utility of
covariance measurements by constructing a plausible scenario in which
a measurement of the covariance of these observables can resolve
ambiguity in individual HBT and antiproton measurements.  After
explaining the basic scenario, we discuss how antiproton production
and HBT radii become correlated in the framework of a thermal model of
hadron production.  We then develop a Monte Carlo code to provide a
realistic simulation of our scenario in the context of STAR.

Measurements of the covariance of distinct signals are useful when
event averaging hides otherwise-strong signatures of a new event
class.  Suppose that there are two event classes -- ``plasma'' and
``hadronic.''  Further assume that the mean antiproton rapidity
density $\overline{N}\equiv d{\overline n}/dy$ is different in each
class, so that the signal is truly strong. Averaging over events
yields
\begin{equation}
\langle \overline{N}\rangle \equiv \sum_i \overline{N}_i =
f \overline{N}(q) + (1-f)\overline{N}(h),
\label{eq:i1}\end{equation}
where the $f$ is the probability that the $i^{\rm th}$ collision forms
a plasma and ${\overline N}(q,h)$ are the average values for the
plasma and hadronic classes respectively.  The event-averaged
$\langle\overline{N}\rangle$ is a smooth function of centrality and
beam energy, because $f$ is a continuous function of the collision
geometry and energy deposition. Smooth data sets, such as those in figs. 1a 
and 1b, are subject to broad interpretation. 

\begin{figure} 
\epsfxsize=4.5in
\centerline{\epsffile{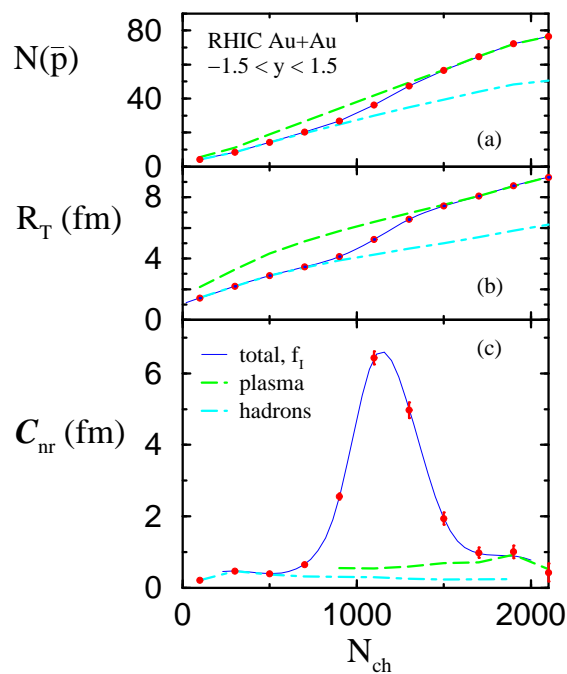}}
\caption[]{
  Antiproton multiplicity (a), pion HBT transverse radius (b) and
  their covariance (c) as functions of the charged particle
  multiplicity. Results are obtained from simulations below for the
  scenario I, cf. eq.  (\ref{eq:pangloss}). Multiplicities
  are computed for the STAR acceptance.  }\end{figure}
The correlation of ${\overline N}$ with the HBT transverse radius $R_T$ 
can betray the existence of the new event class. This correlation is
characterized by the covariance:
\begin{equation}
{\cal C}_{nr}\equiv 
\sum_i (\overline{N}_i - \langle \overline{N}\rangle)
(R_{Ti} - \langle R_T\rangle),
\label{eq:i3}\end{equation}
where $R_T$ is the transverse radius measured in event-by-event 
pion interferometry.  We find 
\begin{equation}
\Delta{\cal C}_{nr} = 
f(1-f)\{\overline{N}(q) -\overline{N}(h)\}\{R_T(q) - R_T(h)\},
\label{eq:i4}\end{equation}
where ${\cal C} \equiv f{\cal C}(q) + (1-f) {\cal C}(h) + \Delta{\cal
  C}$ and the covariance for each class is ${\cal C}({q,h})$.  The
term $\Delta{\cal C}$ depends on the hidden shift in $\overline{N}$
and $R_T$.  In the best of all possible worlds, $f$ will change from
zero to unity for a single target-projectile combination as the impact
parameter $b$ is varied.  This variation can introduce a peak as shown
in fig.~1c, provided that both class-averages $R_T(q)$ and
$\overline{N}(q)$ truly exceed the hadronic values, as predicted. More
likely, one may come upon a low impact-parameter region where $f$
begins to rise, or a peripheral region where $f$ begins to fall.

We point out that the variances $\sigma_{\overline{N}}^{2}$ and
$\sigma_{R_T}^{2}$ can exhibit similar behavior in the presence of two
event classes.  For antibaryons, we find $\langle
\Delta\overline{N}^2\rangle = f\sigma_{\overline{N}}^{2}(q) +
(1-f)\sigma_{\overline{N}}^{2}(h) + \Delta\sigma_{\overline{N}}^2$,
where
\begin{equation}
\Delta{\sigma_{\overline{N}}}^2 =
f(1-f)\{\overline{N}(q) -\overline{N}(h)\}^2,
\label{eq:i2}\end{equation}
and $\sigma_{\overline{N}}^{q,h}$ are the standard deviations for each
class; the result for $R_T$ is similar.  The variance
$\sigma_{\overline{N}}^{2}$ can be particularly interesting in the
context of recent work by Stephanov, Rajagopal and Shuryak
\cite{Krishna}; we address that point elsewhere \cite{sg,gp2}.

A measurement of (\ref{eq:i3}) is built upon an event-by-event
analysis on which we now comment.  We expect anywhere from 50 to 100
antiprotons per event in the STAR acceptance, and possibly more
\cite{Stachel}. Such a number is perfectly adequate for event-by-event
analysis. Since our intent is to use the antiproton yield as a proxy
for an antibaryon measurement, we are not concerned if the
measurements contain a contribution from antilambdas, as found at the
AGS \cite{pbarData}.  On the other hand, the HBT part of the
measurement is challenging because it involves an event-by-event
two-pion correlation analysis.  In an HBT analysis, one compares the
measured identical-pion correlation function $C_{\pi\pi}$ to a
Gaussian parameterization. Radius parameters are typically obtained by
a three dimensional fit to high-statistics data (see \cite{pratt} for
details).  The hadronization of the plasma affects these
spacetime-dependent variables directly by delaying the freezeout of
the system \cite{PrattBertsch}.  While such a three-dimensional
analysis may not be practical for single RHIC events, the feasibility
of one dimensional analyses has been studied by the STAR experiment
\cite{sanjeev}.  One can extract the transverse radius by comparing
\begin{equation}
C_{\pi\pi}(q_T) = 1 + {\rm e}^{-q_T^2 R_T^2} 
\label{eq:gauss1d}\end{equation}
to data, where $q_T$ is the difference in the pions' transverse
momenta.  Alternatively, Heinz and Wiedemann suggest that it may be
easier to extract event-by-event parameters using $R_T^{-2} \propto
\int d^2q_T q_T^2(C(q_T)-1)$ \cite{hw}.  The difference between these
definitions is immaterial to our discussion.  However, we point out
that $R_T$, which essentially measures the transverse system size, is
not an ideal plasma probe.  Nevertheless, we find a strong change in
${\cal C}_{nr}$ for a $50\%$ increase in $R_T(q)$ due to the
transverse growth of the longer-lived plasma system.

\section{Fluctuations near equilibrium}

To establish the mechanisms that drive the correlations and
fluctuations in antibaryon production and pion HBT, we employ the
idealized but standard Bjorken hydrodynamic framework that describes
particle production near rapidity $y=0$.  We assume that matter in
this region is in local thermal equilibrium, with an average local
temperature $T$, entropy density $s$ and net baryon density $\rho_B$
that vary only with proper time $\tau$.  The transverse area $\cal A$
is initially determined by the overlap of the colliding nuclei.  One
can define a comoving volume for matter in the central region, $V
\equiv S/s = {\cal A} \tau$.  This volume grows from a formation time
$\tau_0$ to freezeout at $\tau_F$, so that the entropy per unit
rapidity for all hadrons $S$ is $\tau$ independent.  The total
rapidity density of hadrons $N_{\rm tot}$ is nearly constant, because
$N_{\rm tot} \propto S$ for a system dominated by light hadron species
with masses $\ll T$.  Baryon current conservation implies that the net
baryon rapidity density $N_B \equiv dn_B/dy = {\cal
  A}\rho_B(\tau)\tau$ is independent of proper time.  Mean rapidity
densities of individual species, such as antibaryons $\overline N$ and
baryons $N$, vary with $\tau$.  Observe that this model is distinct
from global equilibrium models used by many groups.  The key
distinction is that here, we focus on thermodynamic quantities local
to the central region.

In local equilibrium, the number of antibaryons at midrapidity and
the source size fluctuate depending on the fluctuation of the state
variables in this region. We consider an ensemble of collisions at
fixed impact parameter in which $V$, $T$ and $N_B$ fluctuate
\cite{LL}.  We assume that these variables are independently
established near $y = 0$ in the preequilibrium evolution, so that they
are statistically independent. The comoving volume and temperature
fluctuate because the initial number of mesons and the initial energy
per meson vary from event to event.  The net baryon number fluctuates because
the central region can exchange baryon current with the rest of the
collision volume for fixed $V$ and $T$, i.e.~the central region is
held at constant baryon chemical potential.  We remark that this
ensemble differs marginally from the familiar Grand Canonical
ensemble, in which the total hadron energy fluctuates and $T$ is held
fixed.  The differences are negligible in the regime 
$N_B \ll N_{\rm tot}$ that we consider.

The fluctuations of a quantity $X$ are characterized by the variance
$\sigma_{X}^2 = \langle\Delta X^2\rangle$ for $\Delta X \equiv X
-\langle X\rangle$. The fluctuations of $V$ satisfy
\begin{equation}
{{\sigma_V^2}\over {V^2}} = {{\sigma_{\rm tot}^{2}}\over {N_{\rm tot}}} 
\sim N_{\rm tot}^{-1},
\label{eq:eq2a}\end{equation}
where $N_{\rm tot}$ is the total number of hadrons.
Thermal fluctuations satisfy \cite{LL}:
\begin{equation}
\sigma_T^2 \approx T^2C_v^{-1}\approx
{T^2\over {12}}~{{\sigma_V^2}\over {V^2}}
\label{eq:eq2b}\end{equation}
for a system of a heat capacity $C_v$, where the second equality
follows for an ideal gas of (mostly) massless hadrons in which $N_B
\ll N_{\rm tot}$ and Fermi and Bose statistics can be neglected.  The
variance of the net baryon number at constant temperature and volume
is
\begin{equation}
\sigma_{B}^2 \equiv \langle \Delta N_B^2\rangle_{{}_{TV}}
= T\partial N_B/\partial \mu_B 
\label{eq:eq3}\end{equation}
by straightforward extension of ref.~\cite{LL}.  For an ideal 
gas, ${\overline N} = Vf(T)\exp (-\mu_B/T)$ and $N_B = 2Vf(T)\; {\rm
  sinh}(\mu_B/T)$ (we need not specify $f$), implying that
\begin{equation}
\sigma_{B}^2 
=  N + \overline{N},
\label{eq:eq4}\end{equation}
neglecting small corrections from Fermi statistics. 

We comment that Stephanov, Rajagopal and Shuryak \cite{Krishna} have
discussed thermal fluctuations of observables at a critical end point,
focusing on the effect of the divergence of $C_V$ on (\ref{eq:eq2b}).
We point out that critical fluctuations also affect (\ref{eq:eq3}),
because $\partial N_B/\partial \mu_B = -\partial^2 \Omega/\partial
\mu^2$ diverges at that point, where $\Omega$ is the free energy.
The antibaryon fluctuations that we discuss shortly would reflect this
critical behavior, c.f. (\ref{eq:eq4c}) below.  Here, we assume that
$T$ and $\mu_B$ are sufficiently far from that point that these
effects are negligible.

The extent to which the number of antibaryons fluctuates depends on
whether or not chemical equilibrium holds with respect to the
reactions that change the number of baryon-antibaryon pairs, e.g.
$N{\overline N} \rightleftharpoons$ mesons.  We do not expect chemical
equilibrium for these processes unless a plasma is formed \cite{ggpv}.
Suppose then that the numbers of baryons and antibaryons are
individually conserved.  For constant $T$ and $V$, ref.~\cite{LL}
implies that the antibaryons essentially follow Poisson statistics,
\begin{equation}
\langle \Delta \overline{N}^2\rangle_{{}_{TV}}
= \overline{N}, \,\,\,\,\,\,\,\,\,\, 
\,\,\,\,\,\,\,\,\,\, {\rm no~chemical~equilibrium}
\label{eq:eq4a}\end{equation}
with a similar expression for protons.  Observe that HIJING and similar 
event generators exhibit exactly the same Poisson-like behavior (such 
behavior is built in to these models).

Chemical equilibrium couples fluctuations of the baryons to those of
the antibaryons, reducing the relative variance compared to
(\ref{eq:eq4a}).  In this case, a small change in the number of
antibaryons requires the net baryon number change:
\begin{equation}
\Delta \overline{N} = 
\left({{\partial {\overline N}/\partial \mu_B}\over 
{\partial N_B/\partial \mu_B}}\right)
\Delta N_B.
\label{eq:eq4b}\end{equation}
The variance in antibaryons is then
\begin{equation}
\langle\Delta \overline{N}^2\rangle_{{}_{TV}} =
T{{(\partial {\overline N}/\partial \mu_B)^2}\over 
{\partial N_B/\partial \mu_B}}
= \overline{N}^2(N+\overline{N})^{-1},
\label{eq:eq4c}\end{equation}
where we have used (\ref{eq:eq4}).  Note that we recover (\ref{eq:eq4a}) in 
the limit $N \ll \overline{N}$, since baryon conservation keeps 
$\overline{N}$ constant.

Volume and thermal fluctuations can also cause the number of antibaryons 
to fluctuate. We write:
\begin{equation}
\Delta \overline{N} = 
{{{\overline N}\Delta V}\over{V}} + 
\left({{\partial {\overline N}}\over {\partial T}}\right)_{{}_{N_B}}
\!\!\!\!\Delta T
+ 
\left({{\partial {\overline N}}\over {\partial N_B}}\right)_{{}_{T}}
\Delta N_B.
\label{eq:eq5}\end{equation}
The third contribution represents the constant T and V results
discussed earlier.  The second term,
\begin{equation}
\left({{\partial {\overline N}}\over {\partial T}}\right)_{{}_{N_B}}
= {{\partial {\overline N}}\over {\partial T}}
- \left({{\partial {\overline N}}\over {\partial \mu_B}}\right)
{{\partial N_B/\partial T}\over {\partial N_B/\mu_B}},
\label{eq:eq5a}\end{equation}
allows thermal fluctuations to change the antibaryon population by
making pairs in chemical equilibrium; it strictly vanishes otherwise.
We will see that the covariance (\ref{eq:i3}) arises from the 
first term.

For an ideal gas in chemical equilibrium,
\begin{equation}
{{\Delta \overline{N}}\over {\overline N}} = 
{{\Delta V}\over{V}} + 
{{2\epsilon N}\over {N +\overline{N}}}{{\Delta T}\over T}
- {{\Delta N_B}\over {N+\overline{N}}}.
\label{eq:eq6}\end{equation}
where $\epsilon = E_{\overline{N}}/\overline{N}T \approx m/T + 3/2$ is
the energy per antibaryon per unit temperature. 
We then obtain:
\begin{equation}
{{\sigma_{\overline N}^2}\over {\overline{N}^2}}
=
{{\sigma_V^2}\over{V^2}} + 
\epsilon^2\left({{2\overline{N}}\over {N+\overline{N}}}\right)^2
{{\sigma_T^2}\over {T^2}}
+ {{\sigma_B^2}\over {(N+\overline{N})^2}}.
\label{eq:eq8a}\end{equation}
The third term, i.e. (\ref{eq:eq4c}), dominates the antibaryon
fluctuations when $N_{\rm tot} \gg N + \overline{N}$, as eqs.
(\ref{eq:eq2a},\ref{eq:eq2b}) suggest. 
For $\overline{N}\gg N$ we obtain the generalization of (\ref{eq:eq4a}),
\begin{equation}
{{\sigma_{\overline N}^2}\over {\overline{N}^2}}
=
{{\sigma_V^2}\over{V^2}} 
+ {1\over {\overline{N}}}.
\,\,\,\,\,\,\,\,\,\, {\rm no~chemical~equilibrium}
\label{eq:eq8b}\end{equation}
The underlying volume
fluctuations follow (\ref{eq:eq2a}) provided that thermal equilibrium
holds.

We expect volume, thermal and baryon-number contributions to be
generally comparable at RHIC in thermal equilibrium.  The sum of the
volume and thermal terms is roughly $1 + \epsilon^2/12\approx 7$ for
$T = 140$~MeV and $m = 938$~MeV, implying a contribution to
$(\sigma_{\overline N} /\overline{N})^2$ of $\sim 0.7\%$ for $N_{\rm
  tot} \sim 10^3$ as expected in Au+Au collisions. Baryon density
fluctuations contribute $\sim 1.3\%$ to the variance for $\overline{N}
\approx N\approx 40$, yielding a total $(\sigma_{\overline
  N}/\overline{N})^2 \sim 2\%$. Observe that either the chemical
nonequilibrium result (\ref{eq:eq4a}) or HIJING would give a similar
value $\sim 2.5\%$.

We expect the covariance (\ref{eq:i3}) to be determined primarily by
volume fluctuations, as we see by computing the related covariance of
$\overline{N}$ and $V$,
\begin{equation}
{\cal C}_{nv}\equiv \langle\Delta \overline{N} \Delta V \rangle = \overline{N}
\sigma_V^2/V.
\label{eq:eq9}\end{equation}
The large contributions to the antibaryon variance (\ref{eq:eq8a})
from net baryon number and temperature fluctuations are absent in
(\ref{eq:eq9}). A very important consequence of this result is that
this expression is valid whether or not chemical equilibrium holds.

To relate the volume fluctuations to fluctuations in the HBT radius,
we observe that variations in HBT transverse radius are driven
primarily by the geometrical effective radius $\sim {\cal A}^{1/2}$
and, secondarily, by flow \cite{HeinzJacak}. By analogy with
(\ref{eq:eq5}), we make the phenomenological ansatz:
\begin{equation}
\Delta R_T/R_T = \kappa\Delta V/2V + \lambda\Delta T/T,
\label{eq:eq9b}\end{equation}
where the parameters $\kappa = 2\partial \log R_T/\partial\log V$ and
$\lambda = \partial \log R_T/\partial \log T$ generally must be
determined by hydrodynamic calculations.  We take the
relative concentrations $N/N_{\rm tot}$ and $\overline{N}/N_{\rm tot}$
to be small enough that the baryons have a negligible effect on the
flow at RHIC.  The covariance of $R_T$ and $\overline N$ is then
\begin{equation}
{\cal C}_{nr} \equiv
\langle\Delta {\overline N}\Delta R_T\rangle = 
\overline{N}R_T\left(\kappa {{\sigma_V^2}\over {2V^2}} 
+ \epsilon\lambda{{\sigma_T^2}\over {T^2}}\right),
\label{eq:eq10}\end{equation}
while the fluctuations in $R_T$ itself satisfy
\begin{equation}
\sigma_{R_T}^2/R_T^2 \equiv
\kappa^2 \sigma_V^2/4V^2 + \lambda^2\sigma_T^2/T^2.
\label{eq:eq11}\end{equation}
In contrast to (\ref{eq:eq9}), there is no first-principles
justification for (\ref{eq:eq10}, \ref{eq:eq11}) -- flow
effects are outside the reach of our thermodynamic approach.

In the absence of transverse flow, $R_T$ is essentially the geometric
transverse radius \cite{HeinzJacak}, so that flow contributions to
(\ref{eq:eq10}) and (\ref{eq:eq11}) must vanish. In keeping with our
Bjorken-like scenario, we therefore take
\begin{equation}
\kappa = 1 \,\,\,\,\,{\rm and}\,\,\,\,\, \lambda = 0.
\label{eq:eq11c}\end{equation}
To see that this approximation is reasonable, we use a parameterization
of the Yano-Koonin-Podgoretskii radius $R_T(M_T)$, eq.~(47) of
ref.~\cite{HeinzJacak}, which is closely related to our $R_T$.  For
that parameterization, $\kappa$ is strictly unity.  We estimate
$\lambda\sim 0.3$ for a plausible mean transverse fluid rapidity of
$\eta_f=0.6$.  This result is in keeping with findings at SPS energy,
where NA49 finds that $R_T(M_T)$ decreases by $\sim 20\%$ as $m_T$ is
varied from zero to $\approx 0.45$~MeV, corresponding to $\lambda \sim
0.2$. One can improve our estimate by introducing stochastic noise
and dissipation into a hydrodynamic model and studying the consequent
fluctuation of $R_T$.

We comment that by using the comoving volume $V$ for the Bjorken
scaling expansion, we can describe the local equilibrium fluctuations
in our evolving system as if the system were stationary.  However, for
more general local equilibrium systems flowing in three dimensions,
i.e., real heavy ion systems, we must obtain the variance and
covariance from the local hydrodynamic fluctuations \cite{LL2}.  To
see how inhomogeneity can affect our estimates, consider temperature
fluctuations that locally satisfy $\langle \Delta T({\vec x})\Delta
T({\vec x}^{\prime})\rangle = [T(\vec{x})^{2}/n(\vec{x})c_v(\vec
x)]\delta(\vec{x} -\vec{x}^{\prime})$, where $n$ is the total hadron
density and $c_v$ the specific heat.  Let us assume that longitudinal
Bjorken scaling holds but that the system is inhomogeneous in the
transverse direction $\vec r$ -- the simplest possible extension of
our idealized model.  Experimental measurements of the average
transverse momentum roughly yield the volume average $T \propto \int
d^{2}r n(\vec{r}) T(\vec{r})$ in each event \cite{Krishna}. We can
then use this volume average to construct the event average $\langle
\Delta T^{2}\rangle$.  Using the above correlation function, we see
that this average satisfies (\ref{eq:eq2b}), since local equilibrium
implies that $T(r)^{2}/c_{v}(r)$ scales as $n^{-1}$, neglecting
corrections from Bose and Fermi statistics. In this case and in the
case of rapidity densities, we do not expect transverse inhomogeneity
to introduce large corrections. However, experiments do not strictly
yield the volume average of $T$ but, rather, momentum distributions
from which we extract an average $p_T$ slopes. To treat fluctuations
of $p_T$ slopes, $R_T$ and other such quantities with precision, we
must use hydrodynamic or transport models that include noise and
dissipation that respect the fluctuation-dissipation theorem
\cite{LL2}. No such models exist.

\section{fluctuations in ion collisions}

To describe the two-event-class scenario outlined in the
introduction, we now apply these general results to develop a Monte
Carlo event generator for correlated signals in RHIC collisions. With
this generator we can account for additional sources of fluctuations
in heavy ion collisions as described below.  Schematically, we
generate events as follows. For each event class, the average
$\overline N$, $R_T$ and total rapidity density $N_{\rm tot}$ are
determined at each impact parameter by the collision geometry using
the prescription that we describe shortly.  We then generate values of
$\overline{N}_i$,$R_{Ti}$ and the charged-particle multiplicity for
each event in accord with the variances and covariance calculated
earlier. As possible in the STAR experiment, we use the charged particle
multiplicity to determine the centrality of each event. We then
compute the covariance as a function of multiplicity for two
illustrative two-class scenarios.  In addition, we present the
distribution of simulated events to eliminate some of the abstraction
that often obfuscates statistical analyses.

In discussing thermodynamic fluctuations, we have so far assumed that
collisions occur at a fixed impact parameter $b$, with fluctuations
occurring about well defined mean values fixed by $V$, $T$ and $N_B$.
Additional fluctuations arise in heavy ion experiments, e.g., from the
imperfect experimental knowledge of the impact parameter. In this work,
we incorporate such fluctuations into our description using a Monte
Carlo framework.  However, one can understand how these fluctuations
come about as follows. Events with impact parameters in the range from
$b-\Delta b/2$ to $b+\Delta b/2$ yield average numbers of antibaryons 
that differ by $\approx (\partial {\overline N}/\partial b)\Delta b$. An 
experiment that does not resolve these impact parameters will measure 
a covariance:
\begin{equation}
{\cal C}_{nr} - {\cal C}_{nr}^{\rm eq} \approx
\left({{\partial {\overline N}}\over {\partial b}}\right)
\left({{\partial R_T}\over {\partial b}}\right)
\sigma_b^2,
\label{eq:fluct1}\end{equation}
%
%
%
where $\sigma_b$ is the standard deviation for events in the
unresolved range and ${\cal C}_{nr}^{\rm eq}$ is the average
equilibrium contribution for collisions at $b$.  This centrality
contribution must vanish for central collisions by symmetry.  More
generally, its magnitude depends on how the mean values $\overline N$
and $R_T$ vary with collision geometry and, additionally, how
centrality is measured.  For Au+Au collisions in STAR, we find that
impact parameter fluctuations are typically more important than volume
fluctuations, but less important than the thermal and baryon-number
contributions.  [At an impact parameter $b = 6$~fm, we use
eqs.~(\ref{eq:pbarHG}, \ref{eq:rtHG}) to estimate the centrality
contribution to the relative covariance ${\cal C}_{nr} /{\overline
  N}R_T$, to be $\sim 0.3\%$ compared a thermal contribution of $\sim
0.05\%$.]

To estimate antibaryon production for the purely hadronic event
class, we observe that kinetic theory estimates suggest that it is
unlikely that baryons will reach chemical equilibrium \cite{ggpv,gp2}.
At RHIC, events too peripheral to produce a plasma will yield fewer
antibaryons than required by detailed balance for the reactions
$N\overline{N} \rightleftharpoons$ mesons.  Chemical equilibration
then requires that meson collisions increase the antibaryon
population, but time scales for those processes greatly exceed the
relevant dynamic time scales.  We can therefore expect the hadron
fluid to be far from chemical equilibrium, although thermal
equilibrium may hold.  In this case, the mean rapidity density of
antibaryons is essentially the same as its initial value, since
antibaryon number is now effectively conserved.  Moreover, we use
(\ref{eq:eq8b}) and (\ref{eq:eq9}) to compute the standard deviation
$\sigma_{\overline{N}}(h)$ and covariance $C_{nr}(h)$.  Observe that
the final state is practically indistinguishable from the
entirely-nonequilibrium initial state.  For Au+Au at RHIC we take:
\begin{equation}
\overline{N}(h) \equiv \overline{n}_h{\cal N}(b)/{\cal N}(0);
\label{eq:pbarHG}
\end{equation}
where $\overline{n}_h = 40$, a value consistent with the range of
event generator predictions: HIJING, HIJING/$B{\overline B}$ (its
successor) and RQMD report rapidity densities of 60, 20 and 20
respectively.  We assume the rapidity density varies with impact
parameter in proportion to the number of participants, ${\cal N}(b)$,
in agreement with HIJING calculations at the 1-2 percent level.  

Antiproton production can be very different in a plasma.  A variety of
mechanisms from chiral restoration to disoriented chiral condensate
formation \cite{qgpPbar} can enhance production of baryon-antibaryon
pairs and facilitate equilibration.  Model calculations typically
yield values of the antiproton rapidity densities well in excess of
event generator estimates. For example, ref.~\cite{Stachel} predicts
values of $\overline{N}$ of about 86, almost three times the HIJING
level. For an enhancement at that level to be hidden in the mean
value (\ref{eq:i1}), $f$ would have to be very small. We assume a more
conservative 30\% enhancement over the hadron gas value in a central
collision,
\begin{equation}
\overline{n}_q \approx 26,
\label{eq:pbarQGP}
\end{equation}
and take the same centrality dependence. Importantly, since chemical
equilibration is likely in a plasma, we now use (\ref{eq:eq8a}) to
calculate $\sigma_{\overline{N}}(q)$. On the other hand the covariance
for chemical equilibrium is still given by (\ref{eq:eq9}), so that
$C_{nr}(q) \approx C_{nr}(h)$.

For our HBT estimate, we take a mean transverse radius in the hadronic 
event to be roughly the geometric radius
\begin{equation}
R_T(h) = r_h[{\cal A}(b)/{\cal A}(0)]^{1/2},
\label{eq:rtHG}
\end{equation}
where $r_h = 6$~fm and ${\cal A}(b)$ is the geometrical overlap area
of the colliding nuclei. For comparison, Hardtke and Voloshin
\cite{Hardtke} has used RQMD to find a side radius $R_s\approx 5$~fm
in Au+Au at RHIC and we expect $R_T\sim R_s$. We then use (20) and
(22) to estimate the fluctuations.

Pratt and Bertsch \cite{PrattBertsch} have argued that plasma
formation can increase the pion HBT radius parameters, e.g., if a
nearly-first-order phase transformation leads to the dramatic increase
of the collision-system lifetime and size. However, as noted earlier,
this effect is only dimly reflected in $R_T$. We assume a $50\%$
increase over the hadronic value,
\begin{equation}
r_q\approx 9~{\rm fm}.
\label{eq:rtQGP}
\end{equation}
We expect the intrinsic uncertainty from (\ref{eq:eq11}) to be much
smaller than the experimental uncertainty.

We assume that STAR will measure the multiplicity of charged pions to
select the centrality of each event. We compute the multiplicity for
each event assuming a Gaussian distribution with an average value
$N_{\rm tot}(b) = N_{\rm tot}(0){\cal N}(b)/{\cal N}(0)$ and a
standard deviation $\sigma_{\rm tot} = \sqrt{N_{\rm tot}}$ consistent
with thermal equilibrium. The scale $N_{\rm tot}(0) = 2100$ is
determined by the initial production regardless of event class, in
accord with entropy and energy conservation; the particular value is
taken from a HIJING simulation in the STAR acceptance. We then
generate the antiproton yield and $R_T$ using (6), (7), (12) and (18).
To each $N_{\overline p}\approx \overline{N}/2$ and $R_T$ value we add
a simulated ``experimental'' fluctuation.  The experimental
fluctuation in $R_T$ is distributed with $\Delta R/R \sim 10\%$, in
accord with \cite{sanjeev}, while that of $N_{\overline p}$ assumes an
ad hoc 95\% detection efficiency.

Importantly, these experimental fluctuations do not affect 
the covariance, although they do affect the scatter of events.
Only correlated uncertainties in the $\overline{N}$ and
$R_{T}$ measurements can alter the covariance. Since these two
measurements are very different, we do not anticipate any 
correlated uncertainty (although real experiments are very
complicated!), as long as the measured $\overline{N}$ and
$R_{T}$ truly come from the same event.

To illustrate the effect of two event classes, we present two ad hoc
scenarios for the onset of plasma events in RHIC Au+Au collisions.  We
start with a Pangloss scenario in which plasma forms with certainty in
all but the most peripheral collisions. The large fluctuations in
stopping and energy deposition inherent in peripheral collisions might
plausibly result in an event class in which plasma does not form.  We
take the probability $f(b)$ for plasma events to be
\begin{equation}
f_{\rm I}(b) = \{1 + \exp[(b-b_{0})/\Delta b]\}^{-1},
\label{eq:pangloss}\end{equation}
a form that would apply in the best of all possible worlds.  The
plasma fraction increases from zero to one within a range $\sim
2\Delta b \sim 1$~fm of $b_{0} = 6$~fm.  In figs.~1a and 1b we show
the average $\overline{N}$ and $R_{T}$ derived from 10,000 simulated
events. We see that the average values vary smoothly between hadronic
and plasma model expectations.  In fig.~1c we present the covariance of
$R_{T}$ from $10^{6}$ events (STAR will obtain that many events in
twelve days).  Our simulations clearly establish the behavior of
(\ref{eq:i4}) within the statistical uncertainties. Observe that the
tails of the distribution are described mainly by (\ref{eq:fluct1}),
with volume fluctuations contributing only at the 10-30\% level.
Further note that in the limit as $\Delta b\rightarrow 0$, the width
of the bump in fig.~1c tends to the value $\sigma_{\rm tot}(b_0)$ set
by the multiplicity distribution.
\begin{figure} 
\epsfxsize=4.5in
\centerline{\epsffile{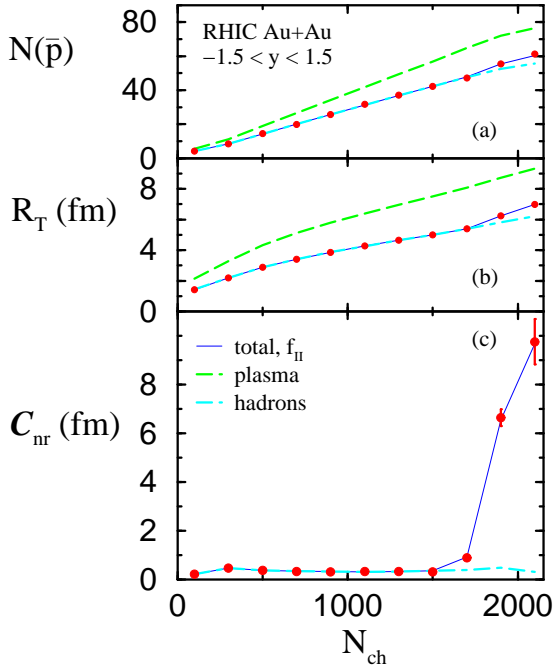}}
\caption[]{
  Antiproton multiplicity (a), pion HBT transverse radius (b) and
  their covariance (c) for scenario II, cf. eq.(\ref{eq:FC}). 
  }\end{figure}

We now consider a more conservative -- but no less ad hoc -- scenario
in which plasma events appear in only a fraction of the most central
collisions. We take:
\begin{equation}
f_{II}(b) = 0.25~[1-(b/b_{0})^{2}]
\label{eq:FC}\end{equation}
for $b<b_{0} = 3$~fm, with $f=0$ otherwise.  We imagine that a region
of $T > T_{C}$ forms in central collisions, and that this region grows
in size as collisions become more central.  The event-averaged Monte
Carlo results in fig.~2a and 2b show no appreciable difference from
hadronic expectations, while the covariance in fig.~2c shows a
striking enhancement.

\begin{figure} 
\epsfxsize=2.5in
\centerline{\epsffile{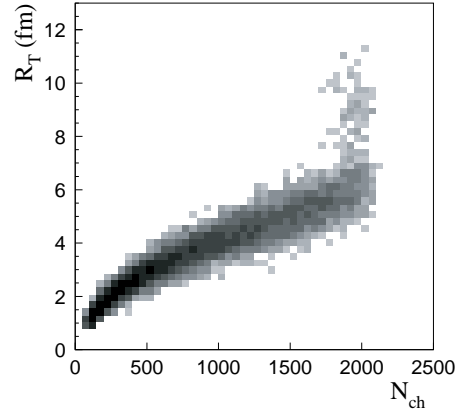}}
\caption[]{
  Distribution of events: transverse radius vs. charged particle
  multiplicity for 10,000 simulated events for scenario II.
  }\end{figure}
\begin{figure} 
\epsfxsize=2.5in
\centerline{\epsffile{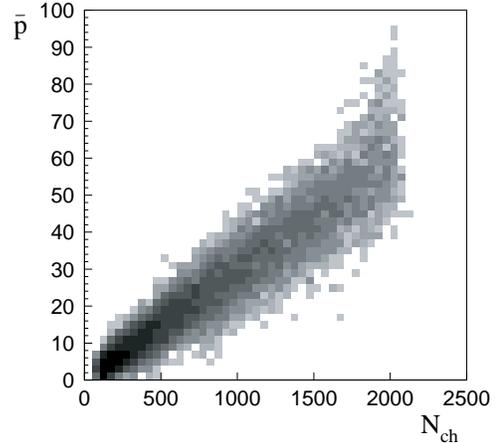}}
\caption[]{
  Distribution of events: antiproton multiplicity vs. charged particle
  multiplicity for 10,000 simulated events for 
  scenario II.  
}\end{figure}
To appreciate how the event distribution gives rise to this behavior,
we show the distribution of events in scatter plots in figs.~3, 4 and
5.  While the antiproton results in fig.~4 are unconvincing, the HBT
events in fig.~3 show a hint of a new population in central
collisions.  The correlated signal from 10,000 events is shown in
fig.~5.  Here, two event classes are clearly distinguishable.  The
scatter plot indicates that we can identify the events with the
largest antiproton yield as anomalous.  Armed with such information,
experimenters can then search for an optimal beam energy and
target-projectile combination to sweep out the coexistence region as
in fig.~1.  Moreover, one can introduce an antiproton trigger to other
experiments to gather a statistically significant sample of plasma
events, e.g., to study hard probes or three dimensional HBT.  We point
out that the covariance can be measured to arbitrary precision by
collecting events and, consequently, is more sensitive to the
appearance of a new event class than is the qualitative scatter plot.
The result in fig.~2c is computed from $10^{6}$ events.
\begin{figure} 
\epsfxsize=2.5in
\centerline{\epsffile{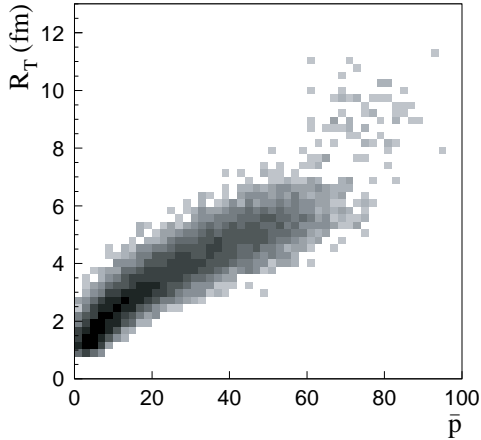}}
\vskip 0.1in
\caption[]{
  Distribution of events: antiproton multiplicity vs. transverse
  radius for 10,000 simulated events for the scenario II.
}\end{figure}

In this work, we have assumed that local equilibrium holds, so that
all variances and covariances are determined by the equation of state
and the conservation laws.  When equilibrium is not in force, there
are very few constraints on correlations.  We study the effect of
incomplete chemical equilibrium on the variances and covariances in
\cite{gp2}.  While such results can be more realistic, they are also
more model dependent. It will eventually be necessary to use models
like RQMD, UrQMD, and VNI to study correlated observables.  Such
models perhaps offer the most realistic description of single particle
observables, but may require modification to address event-by-event
correlations. For example, it has not been demonstrated that
fluctuations in these models respect the fluctuation-dissipation
theorem when one imposes the appropriate boundary conditions and
approximate local equilbrium \cite{LL}.  Consequently, it is not clear
that the dynamical correlations these models produce are realistic.

On the experimental side, it is not clear how to extract
event-by-event HBT radii in practice \cite{sanjeev}. If experimental
fluctuations in HBT radii are too large, it will be necessary to
develop a super-event analysis scheme \cite{jack}.

We have shown that under the right circumstances a correlated analysis of 
observables such as antiproton yield and the pion $R_T$ will augment the 
discovery potential for novel event classes.  In particular, we stress the 
importance of looking for systematic changes in correlations as functions 
of the centrality.  A first study of fluctuations of the pion transverse 
momentum and the $K/\pi$ ratio by the NA49 experiment at the SPS focused on 
central collisions \cite{na49}.  Their study has similar motivations, such 
as testing the degree of equilibration \cite{Mrow,BaymHeiselberg} and 
looking for critical fluctuations \cite{Krishna}.  Centrality studies
would further allow them to turn equilibrium ``off'' or ``on.''   

To demonstrate how correlated signals can help establish a new event
class, we have taken literally many of the predictions of quark gluon
plasma formation. While other assumptions, e.g., chemical equilibrium,
are not necessary to our conclusions, the existence of event classes
with distinctive signals is crucial.  The observables themselves need
not shift abruptly as in a continuum phase transition, but their
values must be markedly different in each class -- correlating
non-signals will not help.

We thank D. Alvarez for HIJING calculations and S. V. Greene, U.
Heinz, T. Miller, R. Pisarski, S. Pratt, J. Rafelski, U. A. Wiedemann
and W. Zajc for discussions. S.G. is grateful to the nuclear theory
group at Brookhaven National Laboratory for hospitality during part of
this work. This work is supported in part by the U.S. DOE grant
DE-FG02-92ER40713.

\end{narrowtext}

\begin{references}

\bibitem{rhic}
STAR Collaboration (J.W. Harris for the collaboration), Nucl. Phys. A566,
277c (1994); (L. Ray for the collaboration), Hyogo 1997, Exciting physics 
with new accelerator facilities, p 219-228 (1997);
Phenix Collaboration (S. Nagamiya for the collaboration) Nucl. Phys. 
A566, 287, (1994); (D.P. Morrison for the collaboration), Nucl. Phys. 
A638, 565 (1998).

\bibitem{qgpPbar}
T. A. DeGrand, Phys. Rev. D30 2001, (1984);
U. Heinz, P.R. Subramanian, W. Greiner, Z. Phys. A318, 247
(1984); P. Koch, B. M\"uller, H. Stocker and W. Greiner, Mod. Phys.
Lett. A3, 737 (1988);
J. Ellis, U. Heinz and H. Kowalski, Phys. Lett. B233, 223 (1989);
J. I. Kapusta and A. M. Srivastava, Phys. Rev. D52, 2977 (1995).

\bibitem{PrattBertsch}
S. Pratt, Phys Rev Lett 53, 1219 (1984); Phys. Rev. D33, 1314 (1986);
G. F. Bertsch, M. Gong, M. Tohyama, Phys Rev C37 1896 (1988);
G. F. Bertsch, G. E. Brown, Phys. Rev. C40 1830 (1989).

\bibitem{pratt}
G. Baym, Acta Phys. Polon. B29 1839 (1998);
S. Pratt, Nucl. Phys. A638, 125c (1998); Phys. Rev. D33
1314 (1986)

\bibitem{HeinzJacak}
U. Heinz and B. V. Jacak, Ann Rev. Nucl. Part. Sci. 
49, 1999; nucl-th/990202.

\bibitem{pbarTheory}
A. Jahns, H. Stoecker, W. Greiner and H. Sorge, Phys. Rev. Lett.
68 2895, (1992); A. Jahns, C. Spieles, R. Mattiello, H. Stoecker, W.
Greiner and H. Sorge, Phys. Lett. B308 11, (1993); Erratum-ibid. B314,
482, (1993);
S. H. Kahana, Y. Pang, T. Schlagel, C. B. Dover, Phys. Rev. C47
1356, (1993); Y. Pang, D.E. Kahana, S.H. Kahana, H. Crawford, Phys.
Rev. Lett. 78, 3418 (1997).


\bibitem{pbarData}
M. J. Bennett et al. (E878 Collab.) Phys. Rev. C56 1521, (1997);
Y. Akiba et al. (E866 Collab.), Nucl. Phys. A610, 139c, (1996);
T. A. Armstrong et al. (E864 Collab.) Phys. Rev. Lett. 79, 3351
(1997); J. Nagle, Yale University Ph. D. thesis (1997); B. A. Cole,
private communications.

\bibitem{Krishna}
M. Stephanov, K. Rajagopal and E. Shuryak, Phys Rev Lett 81, 4816 (1998);
hep-ph/9903292.

\bibitem{sg} S. Gavin, nucl-th/9908070.

\bibitem{gp2} S. Gavin and C. Pruneau, nucl-th/9907040.

\bibitem{Stachel}
P. Braun-Munzinger and J. Stachel, Nucl. Phys. A606, 320 (1996).

\bibitem{sanjeev}
S. Pandey, STAR SVT Proposal; private communication.

\bibitem{hw}
U. A. Wiedemann and U. Heinz, Phys. Rev. C56, 610 (1997) 610.

\bibitem{LL}
E. M. Lifshitz and L. O. Pitaevskiii, Statistical Physics Part 1
(Pergamon, 1980), sec. 115.

\bibitem{ggpv}
S. Gavin, M. Gyulassy, M. Pl\"umer and Venugopalan, Phys. Lett.
234B, 175 (1990).

\bibitem{LL2}
E. M. Lifshitz and L. O. Pitaevskiii, Statistical Physics Part 2
(Pergamon, 1980), sec. 88.

\bibitem{na49}
G. Roland, Nucl. Phys. A638, 125c (1998). 

\bibitem{Mrow} S. Mrowczynski, Phys. Lett. B430, 9 (1998); B439, 6
  (1998); nucl-th/9901078, Phys. Lett. B, in press; M. Gazdzicki,
  Euro. Phys. J. C8, 131, (1999).

\bibitem{BaymHeiselberg} G. Baym and H. Heiselberg, nucl-th/9905022.

\bibitem{Hardtke}
D. Hardtke, STAR Note 384; D. Hardtke and S. Voloshin, nucl-th/9906033.

\bibitem{jack}
J. Sandweiss, private communication.


\end{references}
\end{document}